# Investigation on data fusion of sun-induced chlorophyll fluorescence and reflectance for photosynthetic capacity of rice

Yu-an Zhou[1, 2], Li Zhai[1, 2], Weijun Zhou[3], Ji Zhou[4], Haiyan Cen[1, 2, *]

[1] *College of Biosystems Engineering and Food Science, and State Key Laboratory of Modern Optical Instrumentation, Zhejiang University, Hangzhou, 310058, China*

[2] *Key Laboratory of Spectroscopy Sensing, Ministry of Agriculture, P. R. China Hangzhou 310058, China*

[3] *College of Agriculture and Biotechnology, Zhejiang University, Hangzhou 310058, China*

[4] *Cambridge Crop Research, National Institute of Agricultural Botany (NIAB), Cambridge CB3 0LE, UK.*

**Abstract:** Exploring the potential of photosynthetic capacity is a promising avenue to increase crop yield. However, high-throughput phenotyping of crop photosynthesis is still a labor-intensive task for large-scale genetic screening. Leaf reflectance has long been used to evaluate the leaf biochemical traits, while the retrieval of sun-induced chlorophyll fluorescence (SIF) was also explored in crop phenotyping. By taking the advantage of capturing two reflectance and SIF signals simultaneously, it is hypothesized that the accuracy of assessing photosynthetic capability could be improved. This study aims to investigate the feasibility of estimating two important photosynthetic traits, including the maximum carboxylation capacity ($V_{cmax}$) and the maximum electron transport rate ($J_{max}$), from in-situ leaf reflectance and SIF spectra. Reflectance, SIF, chlorophyll and carotenoids content per area, and $CO_2$ response curves of 149 leaf samples from two rice cultivars were measured. After removing the bands with low signal-to-noise ratio, the SIF spectra used in this study were 665-845 nm with reflectance spectra of 400-2,400 nm. Two different levels of data fusion (raw data fusion and feature fusion) fusing three different types of spectral features were adopted to predict the photosynthetic capability of rice. Competitive adaptive reweighted sampling (CARS) were used to extract features from reflectance and SIF signals, and partial least squares regression (PLSR) was then employed to build regression models based on the extracted features. Results from the single data source showed that $Jmax_{,25}$ and $V_{max,25}$ could be estimated from leaf reflectance or SIF spectra (e.g. $R^2 = 0.56$ for reflectance and $V_{cmax,25}$, $R^2 = 0.63$ for reflectance and $Jmax_{,25}$). The analysis of model variables indicated that the mechanism of SIF to evaluate photosynthetic ability may be different from the mechanism of albedo to evaluate photosynthetic ability. Compared with the single data source model, the measurement-level data fusion model achieved $Jmax_{,25}$ $R^2$ of 0.60 and $Vcmax_{,25}$ $R^2$ of 0.68. RMSE values were reduced by 23% and 21% compared with the best results of single data source. Although mid-level fusion did not improve $R^2$ compared to low-level fusion, its value of RMSE was smaller. In the evaluation of $Jmax$ and $Vc,max$, the $R^2$ of the decision-level data fusion model increased by 21% and 20%, respectively, and the RMSE decreased by 12% and 17% compared with the optimal single data model. And in the prediction of $Jmax$ and $Vc,max$, the decision strategies were often different. The overall results showed that the SIF spectra can effectively predict the photosynthetic capacity of rice, and data fusion strategy improved the model accuracy of evaluating two photosynthetic parameters, which has the potential for high-throughput phenotyping of crop photosynthetic capacity.

# Highlight

1. Continuous SIF spectra can effectively predict the photosynthetic ability of rice.
2. SIF spectra and reflectance spectra have different mechanisms for evaluating vegetation photosynthetic capacity.
3. The continuous SIF spectra predicted $J_{max}$ better, while the reflectance predicted $V_{c,max}$ better.
4. Data fusion helped to compensate for the shortcomings of different data sources and achieved better results.

# 1. Introduction

The United Nations Food and Agricultural Organizations has forecast that by 2050, the world's demand of primary foodstuffs will great increase (Long 2014). However, only by maintaining the current historical production growth rate, rice will only meet 72% of the expected global demand (Ray et al. 2013). Therefore, to meet this huge challenge, it is very important to improve the efficiency and productivity of crop photosynthesis (Ray et al. 2013). Maximum electron transport rate ($Jmax$) and maximum Rubisco carboxylation ($Vc,max$) always are used as indicators of photosynthetic capacity, and improving photosynthetic productivity for increased global crop yields requires techniques to quantify these key parameters (Meacham-Hensold et al. 2019). Rubisco, one of the most abundant proteins on earth, catalyzes the photosynthetic fixation of carbon dioxide (Sharwood 2017). Traditional methods rely on leaf sampling and analysis under laboratory conditions or using in-field gas exchange systems to obtain the above parameters (Long and Bernacchi 2003). Although these methods provide a wealth of photosynthetic information, they are costly and time-intensive. In addition, gas exchange measurements require a long time for leaves to adapt to the controlled environment in the chamber(Lawson et al. 2012; Salter et al. 2019); it also requires precise control of the temperature, humidity, $CO_2$ concentration and other parameters inside the leaf chamber(Berry and Goldsmith 2020; Matthews et al. 2017). In particular, it is very difficult to measure in the water environment where rice located (Du et al. 2020). Due to the constraints of traditional technologies such as gas exchange measurement methods, emerging technologies are urgently needed to solve the challenge of how to quickly and accurately quantify $Jmax$ and $Vc,max$.

Phenotyping refers to the measurement of any aspect of plant growth, development and physiology (Hickey et al. 2019). Faster, better plant phenotyping technology provides the possibility to break through the above bottlenecks (Furbank and Tester 2011), and some researchers have used this technology to evaluate the photosynthetic ability of plants. Dechant et al. (2017) measured the $Vc,max$, $Jmax$, LMA (leaf mass per area), Na (nitrogen content per area) and reflectance spectra of 242 leaves from 37 temperate deciduous tree species, using PLSR (partial least squares regression), with the relationship between Na and photosynthetic capacity, obtained models with good accuracy ($R^2$ = 0.64 for $Vc,max$, $R^2$ = 0.70 for $Jmax$). Heckmann et al. (2017) assessed the potential of leaf reflectance spectra to predict parameters of photosynthetic capacity in $C_3$ and $C_4$ crops. Besides, the researchers also evaluated the performance of various machine learning methods and then selected PLSR as the algorithm of choice that yielded the highest predictive power. Silva-Perez et al. (2018) built prediction models for various biochemical and physiological traits based on wheat data of 76 genotypes, in which the correlation coefficients of 0.62 for *Vc,max*, 0.70 for *Jmax*. Based on the *Jmax*, *Vc,max*, [N] (percentage leaf nitrogen) and reflectance spectra of wild type and genetically modified tobacco for two years, Meacham-Hensold et al. (2019) built multiple PLSR models with certain reliability. The models obtained in these researches generally had good accuracy and certain robustness, but most of them establish indirect linked with photosynthesis through parameters such as nitrogen and pigments, or required complex measurement parameters. In addition, reflectance may be affected by leaf structure interference and chlorophyll concentration, making the prediction model invalid in some cases (thick > 300 μm, chlorophyll content > 100 μg $cm^{-2}$) (Van Wittenberghe et al. 2014). The above factors limited the general applicability of the models.

Chlorophyll fluorescence is the endogenous light emitted by the plant itself, and it is one of the most easily captured remote sensing signals along with reflectance. Compared with reflectance spectra, chlorophyll fluorescence is more closely related to the physiological activities of plants, and it provides for the first time a direct measurement related to plant photosynthetic activity (Guan et al. 2016). It has been proved that sun-induced chlorophyll fluorescence (SIF) is directly affected by photosynthesis, so the analysis of SIF is crucial in the research of related fields (Lazar 2015; Stirbet and Govindjee 2011). Several studies have demonstrated that SIF can estimate photosynthesis and decrease the difficulty of modeling photosynthesis by reducing the parameterization burden and associated uncertainties (Han et al. 2022; Tremblay et al. 2011). Han et al. (2022b) found that the product of $SIF_{PSII}$ and the fraction of open PSII reactions $q_L$, was a strong predictor of both *Vc,max* and *Jmax*, although their precise relationships vary somewhat environmental conditions. The study by Fu et al. (2021) proved that the SIF yield on a single measurement day can predict Jmax and Vc,max, but when SIF yield of all measurement days was used, the regression analysis was not statistically significant. However, most researches often only use the chlorophyll fluorescence parameters obtained based on pulse amplitude modulated (PAM) technology or individual bands, and hardly consider the entire SIF spectra. Therefore, exploring the effect and mechanism of SIF spectra in assessing plant photosynthetic capacity is very helpful for the development of plant phenotypes.

Information fusion aims to reveal the benefits of multi-sensor measurements which are expected to outperform single sensors, providing more accurate assessments results. For instance, in addition to biochemical components, the structure of leaves also has an important impact on spectral information. In particular, SIF and reflectance can be measured in situ simultaneously, making data fusion of the two data sources very significant. In the past few decades, the methods for estimating biochemical and physiological parameters from spectral data were mainly based on simple parametric regression algorithms, and the trend of using machine learning and data fusion strategies is increasing in the future (Berger et al. 2022; Berger et al. 2020). Different data fusion strategies can not only help researchers obtain more effective prediction models, but also provide an in-depth understanding of the operating mechanism of the models (Barbedo 2022; Zhou et al. 2020).

In this study, PLSR model and data fusion strategies are used to predict photosynthetic capacity from rice leaves reflectance spectra and leaf SIF spectra over multiple time periods. Specifically, the objectives are to determine (1) whether the single data source spectral models based on PLSR predict photosynthetic capacity of rice, (2) whether the models based on data fusion strategies contribute to the improvement of model accuracy, and (3) whether SIF can predict $J_{max}$ and $V_{c,max}$ independent of leaf nitrogen.

## 2. Materials and methods

### 2.1 Experimental design

In 2022, two cultivars of rice (Huanghuazhan and Xiushui134) were grown in the net room at the China National Rice Research Institute (CNRRI, 30°08′N, 119°93′E). The rice seedlings were first sown in a paddy field, and then part of them was transported into pots in net room in mid-July. 8 kg of soil was used for each pot. In order to obtain as wide a range of data sets as possible, three nitrogen gradients were designed for each cultivar, with 1.7 g, 3.5 g, and 7 g nitrogen fertilized per pot, respectively, and potassium and phosphorus fertilizers were fertilized according to the most suitable amounts. Irrigation was provided as needed to eliminate water limitation throughout growth. In each measurement day, each set of parameters such as reflectance,

SIF, and *A/Ci* curve was measured at the same location on the rice leaf. There was a total of 20 measurement days during the test period, each measuring 5-15 samples. In August, the early ripening HHZ was mainly measured, and in September, the late rice XS134 was mainly measured. Due to the influence of crop growth cycle, planting environment and weather, the number of samples of two cultivars of rice with different nitrogen levels were finally obtained as shown in the table below:

Table 1. The number of leaves under different cultivars and different nitrogen gradient[a].

| Cultivar | N1 | N2 | N3 |
| --- | --- | --- | --- |
| Huanghuazhan (HHZ) | 9 | 17 | 17 |
| Xiushui134 (XS134) | 13 | 42 | 51 |

[a]: Due to factors such as growth cycle and climate influence, the number of leaf samples under different conditions has certain differences.

**2.2 Leaf sun-induced chlorophyll fluorescence and reflectance measurements**

Leaf spectral reflectance was measured in situ from 400 to 2,500 nm using an ASD FieldSpec Pro FR2,500 spectrometer (Analytical Spectral Devices, Boulder, CO, USA), with the spectral resolution of 3 nm in the visible and near-infrared (NIR; 350-1000 nm) and 8 nm in shortwave-infrared (SWIR; 1000-2500 nm). Measurements were made with a leaf chip attached to the fiber optic cable. The device includes a radiometrically calibrated light source that was standardized for relative reflectance (white reference) before each measurement using a spectral on panel (Meacham-Hensold et al. 2019).

The SIF datasets were performed with an ASD FieldSpec Pro FR2,500 spectrometer coupled with the FluoWat leaf clip (Producción por mecanizados villanueva S.L. U, Spain) under natural illumination with clear sky conditions (Jia et al. 2018; Van Wittenberghe et al. 2013). A high-performance low filter (< 650 nm, Producción por mecanizados villanueva S.L. U, Spain) was used to cut off the light above 650 nm in order to obtain the upward and downward sun-induced fluorescence emission. Additional details regarding the data acquisition are provided in Appendix A.

Because there is a negative correlation between SIF and photochemistry under low light unstressed conditions, and a positive correlation under plant stress and high light conditions (Meroni et al. 2009; van der Tol et al. 2009). Therefore, measurements were made on clear sky days with similar irradiance. In each cultivar plot, 3 to 5 leaves were selected that were wide enough to cover the instruments' field of view for measurement. The measurement point was half of the distance from the root of the leaf. Five repeated measurements were made at each point, and their average value was taken as the measurement value. Since ASD FieldSpec uses three sensors (VIS, SWIR, and NIR), smoothing was performed using the Splice Correction command in ViewSpec Pro software (Version 6.2).

Since the intensity of incident light was different on each test day throughout the rice growth stages, the unitless parameter SIF yield could be calculated by normalizing the absorbed incoming photosynthetically active radiance (*APAR*) to avoid rapid change of SIF in the natural environment (Van Wittenberghe et al. 2013; Van Wittenberghe et al. 2015). APAR equals the integration of incoming sun radiance (*I*) in the photosynthetically active radiance (*PAR*) region (400 – 700 nm) multiplied by the fraction of the light absorbed in the *PAR* region of (*fAPAR*) (Equations (1) – (3)). The steady-state fluorescence was normalized by *APAR* to calculate the upward and downward SIF yield (Equations (4) – (5)).

$$PAR = \int_{400}^{700} I \cdot d\lambda \quad (1)$$

$$f\ APAR = \int_{400}^{700} A \cdot d\lambda = \int_{400}^{700} (1 - R - T) \cdot d\lambda \quad (2)$$

$$APAR = f\ APAR \times PAR \quad (3)$$

$$upward\ SIF\ yield = upward\ F/APAR \quad (4)$$

$$downward\ SIF\ yield = downward\ F/APAR \quad (5)$$

where *A* stands leaf absorbance, *R* stands leaf apparent reflectance (contains fluorescence emission), *T* stands leaf apparent transmittance (contains fluorescence emission), *F* stands measured steady-state fluorescence.

## 2.3 Leaf photosynthesis measurements and parameter estimation

*2.3.1 Leaf Jmax and Vc, max measurements*

Photosynthetic (*A*) vs intercellular (*Ci*) response curves were measured on the same leaves as the spectra measurements using a portable leaf gas exchange system (Li-6800, Li-Cor Biosciences, Lincoln, NE, USA). The photosynthetic (*A*) vs light response curves of rice leaves were measured before the experiment started. Therefore, the PAR was set to 1600 μmol m$^{-2}$ s$^{-1}$, which is consistent with the PARsat (saturation irradiance) in previous literature (Wang et al. 2015). The $CO_2$ concentrations were adjusted stepwise over a range of 0 to 1600 μmol mol$^{-1}$ in set increments as follows: 400, 300, 200, 100, 50, 0, 400, 400, 600, 800, 1000, 1200, 1400, 1600. Because of the hot weather, a set point of 30 °C was chosen for most response curves. Leaves were acclimated to chamber conditions for a minimum wait time of 180 s prior to initiating each *A/Ci* curve and a minimum and maximum wait time of 120s and 240s, respectively, was incorporated before triggering each individual measurement. Besides, relative humidity inside the chamber was automatically controlled at 50%. *Jmax* and *Vc,max* were calculated from the measured *A/Ci* curves according to the mechanistic model of photosynthesis (Farquhar et al. 1980). *A/Ci* curves were analyzed by using a curve fitting utility developed by Sharkey et al (Sharkey et al. 2007). The temperature responses of *Jmax* and *Vc,max* were adjusted according to the researches of Wang et al and Bernacchi et al, and the values of the two at 25 °C were calculated (Bernacchi et al. 2001; McMurtrie and Wang 1993).

*2.3.2 leaf chemical analysis*

After field measurements, leaf samples were sent to the laboratory and cut into small discs with a diameter of 8.5 mm. The chemical analysis of the chlorophyll (a + b) and carotenoids content was performed with ethanol (95% v/v) until the samples became colorless when immersed in the solution. The absorbance values (A470, A649, A665) at 470, 649, and 665 nm were measured with a microplate spectrophotometer (Epoch 2, BioTek Inc., USA). Then, the following equations were used to determine the concentrations (μg/mL) of chlorophyll a, chlorophyll b, and (Cona, Conb, and Conxc) (Li et al. 2018):

$$C_{on_a} = 13.95 \times A_{665} - 6.88 \times A_{649} \quad (6)$$

$$C_{on_b} = 24.96 \times A_{649} - 7.32 \times A_{665} \quad (7)$$

$$C_{on_{xc}} = (1000 \times A_{470} - 2.05 \times C_{on_a} - 114.8 \times C_{on_b})/245 \quad (8)$$

The contents (μg/cm²) of *Cab* and *Cxc* were derived using the following equations:

$$C_{ab} = (C_{on_a} + C_{on_b}) \times V / A \quad (9)$$

$$C_{xc} = C_{on_{xc}} \times V / A \quad (10)$$

where *V* is the volume of the centrifuge tube, and the volume in this experiment was 2 ml. *A* represents the area of the leaf, which is the aforementioned small disc with a diameter of 8.5 mm.

*2.3.3 PROSPECT inversion*

Because the small fragments of leaves were very light, the system error was relatively large, and small fragments in situ have been used for chemical measurements of photosynthetic pigments. In this study, the carbon-based constituents (CBC) and nitrogen-based constituents (proteins) of leaves were retrieved by using the published radiative transfer model PROSPECT-PRO (Féret et al. 2021).

**2.4 Data feature extraction and data fusion strategies**

*2.4.1 Competitive adaptive reweighted sampling*

By employing the simple but effective principle "survival of the fittest" in Darwin's Evolution Theory, a novel strategy for selecting an optimal combination of key wavelengths of multi-component spectral data, named competitive adaptive reweighted sampling (CARS), is developed (Li et al. 2009). The core of CARS is to sequentially select wavelengths of different importance from Monte Carlo (MC) sampling runs in an iterative and competitive manner to establish calibration models. CARS can often locate some optimal combination of key wavelengths which can be interpreted as chemical properties of interest (Fan et al. 2011; Xing et al. 2021). In order to obtain the characteristic bands as much as possible, the bands whose occurrence frequency were more than 50% were selected from the 100 cycles as the characteristic bands.

*2.4.2 Measurement-level data fusion and feature-level data fusion*

Data fusion defines protocols to describe how to merge the data from different sources together and give a comprehensive understanding to an object (Zhou et al. 2020). This study used measurement-level data fusion, feature-level data fusion and decision-level data fusion. For the first level of fusion, raw measurements from different tools/methods are simply concatenated as a new dataset for further processing, which is called measurement-level data fusion. Since the value of reflectance is distributed between 0-1, and the value of SIF yield is extremely small, the latter is first normalized and the spliced. The direct combination of raw data also contains noise and redundant information, which affects the accuracy and speed of modeling and calculation. The raw data is processed using feature extraction algorithm, and then the obtained features are used as models' input, which is defined as feature-level data fusion. In this study, the CARS algorithm was used to extract the characteristic wavelengths of reflectance and SIF spectra.

*2.4.3 Majority vote algorithm and decision-level data fusion*

Condensing the diagnostic results of multiple data analysis models to conduct a comprehensive evaluation of the final decision is called decision-level data fusion (Zhou et al. 2020). The majority vote algorithm is used to make final predictions based on the prediction results of a single model (Ballabio et al. 2018; Feng et al. 2020), and is a typical decision-level data fusion strategy. In this study, when the difference between the predicted values of the three data sources was greater than the threshold, the median was taken. When the difference between the predicted values was less than the threshold, the average was taken. At other times, the mean of the two closer numbers is taken. Decision-level data fusion has the potential to reduce the interference of flaws in different decision models, but it also risks losing key information in the original data. Figure 1 showed the flow chart of the three data fusion strategies.

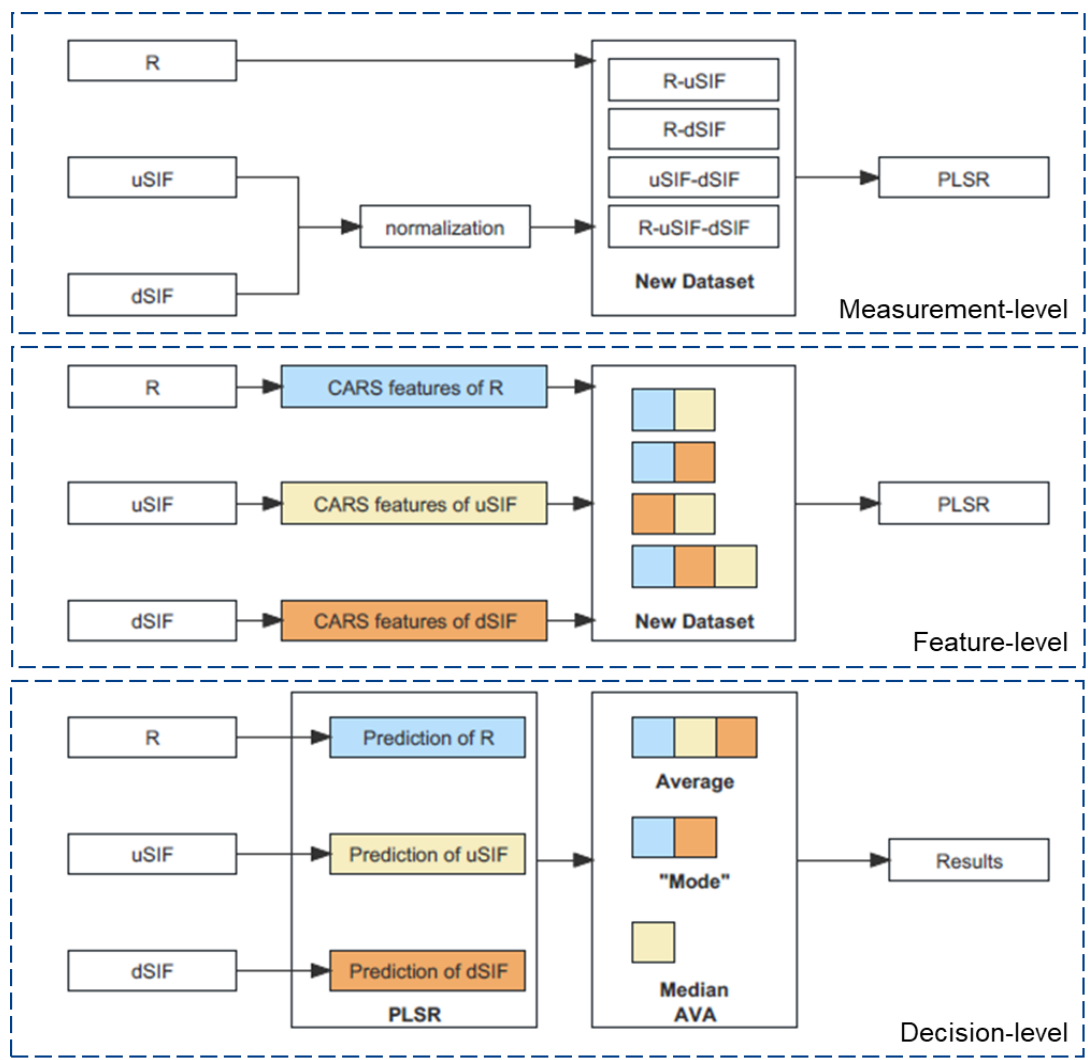

**Fig. 1**. Schematic diagram of the structure of three levels of data fusion. From top to bottom is the data fusion of measurement-level, feature-level and decision-level. R is the abbreviation of reflectance, uSIF is the abbreviation of upward SIF yield, and dSIF is the abbreviation of downward SIF yield. A quoted mode indicates the average of two closer predictions.

**2.5 Models establishment and evaluation**

*2.5.1 Regression models*

Partial least squares regression (PLSR) is a classic multivariate statistical method whose advantage is that the variables and observations increase, and the accuracy of the model will also improve (Wold et al. 2001). PLSR reduces the full-spectrum data into a small set of independent latent variables or factors that are used as new predictors. 10-fold cross-validation was used in this study to calculate and determine the number of factors used in the PLSR model to avoid the problem of overfitting.

Data splitting is a fundamental step for building models with spectral data. The samples are divided into a training set and a test set; herein, the training set is used for model construction, and the test set is used for model validation (Morais et al. 2019). In each modeling process, 2/3 of the dataset was randomly selected as the training set, while the remaining 1/3 of the dataset was used for the test set.

*2.5.2 Data analysis*

First, the mean, maximum and minimum values of each parameter of the measurement dataset were sorted out and compared with previous studies. The variable importance in projection (VIP) algorithm was also used in this study to evaluate the contribution of each band of each data source to the models. The following mathematical metrics are used to measure the accuracy of the models. By comparing the differences between the coefficient of determination ($R^2$) and root mean square error (RMSE) in the predictions, the performances of different spectral indices and multivariate models were evaluated. The higher the $R^2$ and the lower the RMSE are, the better the precision and accuracy provided by the index or model. The $R^2$ value was calculated as follows:

$$R^2 = 1 - \frac{\sum_n (y_n - \hat{y})^2}{\sum_n (y_n - \bar{y})^2} \quad (11)$$

where y, $\hat{y}$ and $\bar{y}$ are the measured, predicted and average values of pigments, respectively, and n is the number of samples.

The RMSE value was calculated as follows:

$$RMSE = \sqrt{\frac{1}{n}\sum_{m=1}^{n}(y_m - \hat{y}_m)^2} \quad (12)$$

where y and $\hat{y}$ are the measured and predicted values of pigments, respectively, and n is the number of samples.

Finally, principal component analysis was used to represent the leaf spectral diversity of crops in different growth stages and different cultivars.

## 3 Results

**3.1 Spectral profiles and distribution of physiological parameters**

After removing outliers, the spectral distribution of all rice samples was plotted, as shown in Figure 1. The measured reflectance spectra, in the band intervals related to chlorophyll content (450-750 nm), mesophyll structure (800-1250 nm), water (1300-2400 nm) and dry matter content (1600-2500 nm), has some variety. Compared with other datasets, the reflectance spectra of this dataset varied less in the visible region (Dechant et al. 2017; Meacham-Hensold et al. 2019). Compared to the upward SIF, the downward SIF has only one peak in normally growing plants due to reabsorption (Buschmann 2007). Although three different nitrogen gradients were designed in this experiment, there was no peak at around 680 nm in the downward SIF spectra, which indicated that the rice planted with a lower nitrogen gradient still had a normal growth environment (Jia et al. 2018; Van Wittenberghe et al. 2015). The specific details were shown in the Figure 2.

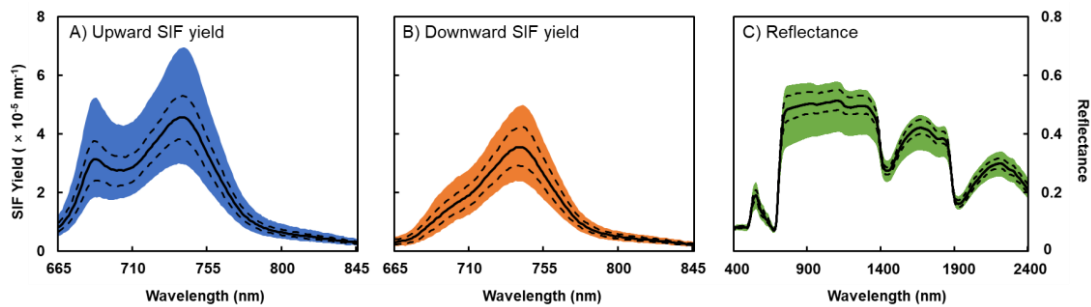

**Fig. 2.** Overview of upward SIF yield (Fig. A), downward SIF yield (Fig. B) and reflectance (Fig. c) spectra measured on 149 individual leaves from 2 cultivars. The average spectra (continuous black line), the first and third quartiles (dashed black lines) and the range of all measured spectra (shaded area) are shown.

In this study, widely distributed *Jmax* values (57.86-141.17 μmol m$^{-2}$ s$^{-1}$) and *Vc,max* values (29.11-101.93 μmol m$^{-2}$ s$^{-1}$) were obtained by calculating *A/Ci* curves through related program tools. In general, both *Jmax* and *Vc,max* had a trend of rising first and then falling, and the change range of Vc,max was larger than that of Jmax. In addition, Figure 3 showed that the average values of $N_1$ and $N_2$ gradients were similar, except that the average value of $N_3$ of Xiushui 134 was significantly higher than that of $N_1$ and $N_2$. The data distribution range of XS134 was wider than that of HHZ, but the average value of the two was not much different. The above rules are applicable to the data distribution of *Jmax* and *Vc,max*. More detailed information can be found in Appendix B.

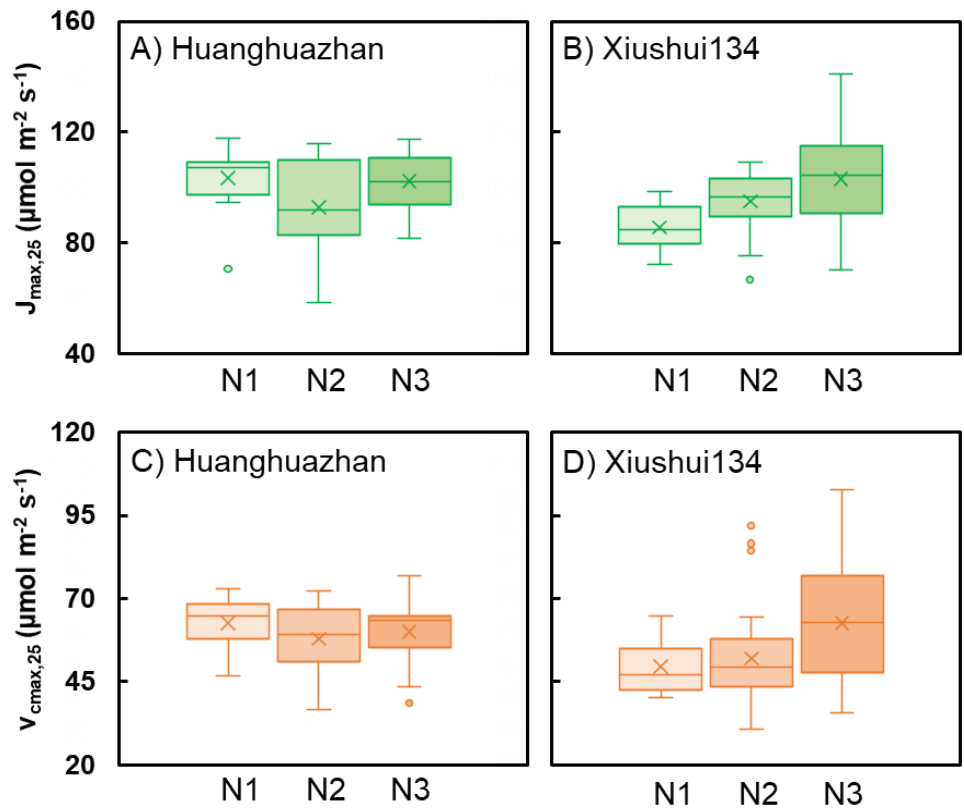

**Fig. 3.** Box plots for *Jmax* (Figure 3.A & Figure 3.B) and *Vc,max* (Figure 3.C & Figure 3.D) calculated from photosynthetic-$CO_2$ response curves for two rice cultivars over 3 nitrogen gradients. The boxes show the interquartile range with the median as solid horizontal line. Whiskers show data outside the interquartile range but within 1.5× the interquartile range. Dots show outliers.

**3.2 Regression models based on single data source**

Both for reflectance and SIF, the PLSR models evaluating *Jmax* and *Vc,max* showed that the single data source models can effectively evaluate the parameters of photosynthetic capacity. the $R^2$ of all models exceeded 0.5, and the RMSE is relatively low compared with previous studies. In addition, the effect on *Jmax* was better than

$Vc,max$. The RMSE of $Jmax$ was less than 12%, while the RMSE of $Vc,max$ was less than 20%. From different data sources, the models based on upward SIF yield and reflectance achieved the best results in the prediction of $Jmax$ and $Vc,max$, respectively.

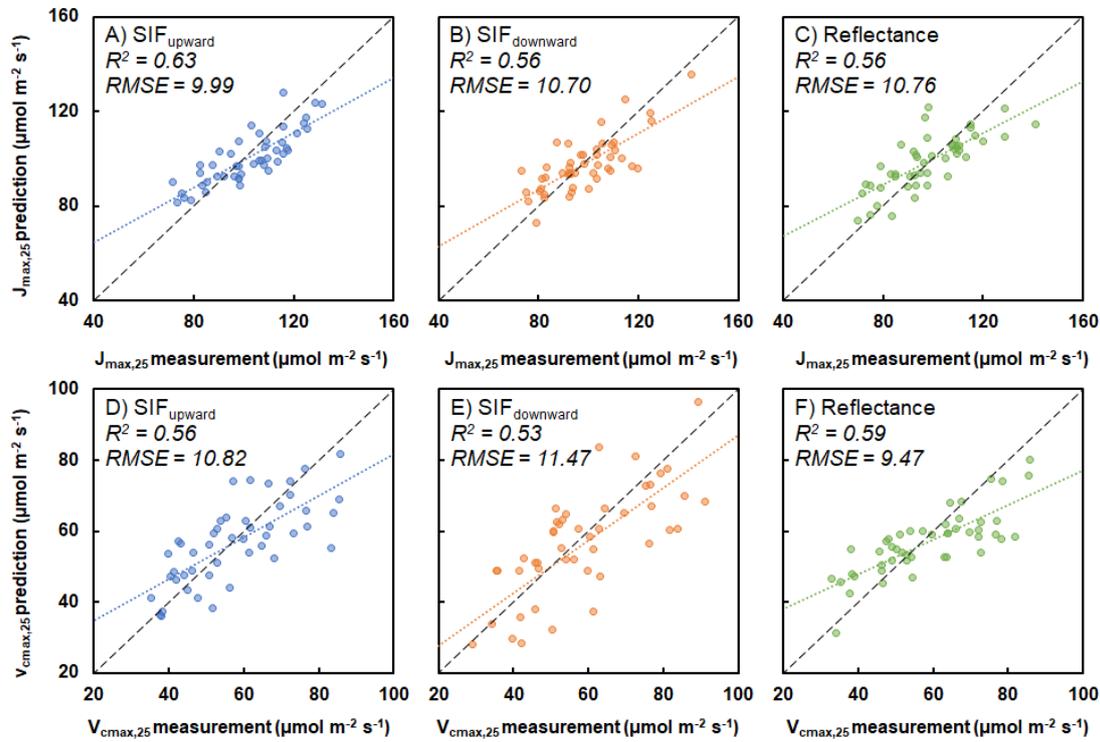

**Fig. 4.** Predicted results from leaf reflectance spectral measurement (400-2400 nm), SIF yield spectral measurement (665-845 nm) vs. measured results by using traditional techniques of *Jmax*,25 (A, B, C) and *Vc,max*,25 (D, E, F). The black dashed line shows the 1:1 line and the statistical results are inset. The colorful dashed lines represent the linear regression fit to the data. AD, BE, CF are the results based on upward SIF yield, downward SIF yield and reflectance, respectively.

### 3.3 Regression models based on data fusion

*3.3.1 Results of measurement-level data fusion*

Since there were three data sources of upward SIF yield (upSIF), downward SIF yield (downSIF), and reflectance (R), in order to understand the effect of data fusion results as much as possible, a total of the four combinations were listed as follows: R-upSIF (this meant the fusion of reflectance spectra and upward SIF yield spectra, and other abbreviations were similar to this fashion), R-downSIF, upSIF-downSIF and R-upSIF-downSIF. From Table 2, it can be found that for the evaluation of *Jmax*, the optimal measurement-level fusion model (upSIF-downSIF) had an $R^2$ improvement of 4.76% and an RMSE reduction of 23.12 % compared to the optimal single data source model (upward SIF yield). For the evaluation of *Vc,max*, the optimal measurement-level fusion model (R-upSIF) improved $R^2$ by 11.86% and reduced RMSE by 17.95% compared to the optimal single data source model (reflectance). The data fusion models of SIF and R were better than their respective single models.

Table 2. The statistical results of different measurement-level data fusion models

| New Dataset | Jmax | | Vc,max | |
| --- | --- | --- | --- | --- |
| | $R^2$ | RMSE | $R^2$ | RMSE |
| R-upSIF | 0.61 | 8.41 | **0.68** | **7.77** |
| R-downSIF | 0.60 | 8.29 | 0.62 | 7.98 |
| upSIF-downSIF | **0.66** | **7.68** | 0.60 | 8.58 |
| R-upSIF-downSIF | 0.60 | 8.44 | 0.66 | 7.52 |

### 3.3.2 Results of feature-level data fusion

The characteristic bands of each data source were extracted according to the multi-cycle CARS algorithm described in the previous chapter. As shown in Figure 3, out of a total of one hundred cycles, only eight reflectance band were selected because they appeared more than 50 times. Correspondingly, many SIF bands have been selected. In fact, the number of upward SIF was more than that of downward SIF.

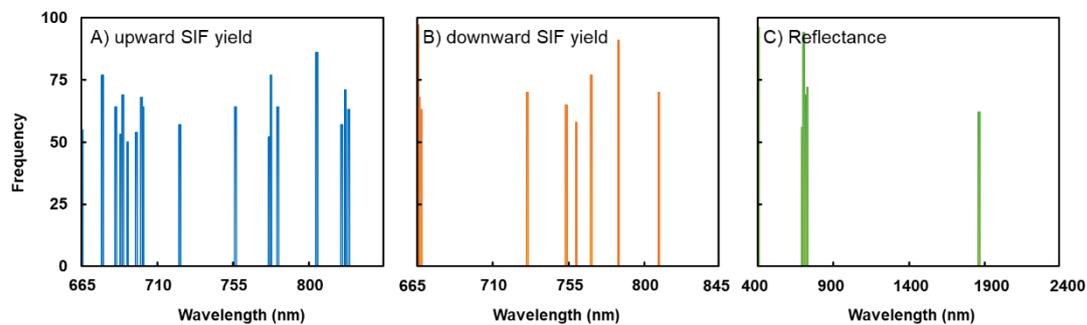

**Fig. 5.** The results obtained by looping through the CARS feature band extraction algorithm 100 times. In order to facilitate the presentation of key points, bands with a frequency of less than 50% have been deleted from the figure.

Similar to measurement-level fusion, four different combination were also used in feature-level fusion. From the results, for the evaluation of *Jmax*, the optimal feature-level fusion model (R-upSIF) reduced $R^2$ by 3.17% and RMSE by 13.71% compared to the optimal single data source model (upward SIF yield). For the evaluation of *Vc,max*, the optimal feature-level fusion model (R-upSIF-downSIF) improved $R^2$ by 6.78% and reduced RMSE by 7.49% compared to the optimal single data source model (reflectance). although the feature-level fusion models did not improve $R^2$ significantly compared to the models based on the single data source, it effectively reduced the

RMSE. Moreover, compared with splicing original spectra, the input of feature-level fusion was greatly reduced.

Table 3. The statistical results of different feature-level data fusion models

| New Dataset | Jmax | | Vc,max | |
|---|---|---|---|---|
| | $R^2$ | RMSE | $R^2$ | RMSE |
| R-upSIF | 0.61 | 8.62 | 0.60 | 9.12 |
| R-downSIF | 0.57 | 8.61 | 0.57 | 9.13 |
| upSIF-downSIF | **0.66** | **7.80** | 0.52 | 9.53 |
| R-upSIF-downSIF | 0.58 | 8.53 | **0.63** | **8.76** |

*3.3.3 Results of decision-level data fusion*

Decision-level fusion achieved very perfect results. For the evaluation of both Jmax and Vc,max, the decision-level fusion model obtained the best $R^2$. But the RMSE was a little higher than all except the single data source model.

Table 4. The statistical results of decision-level data fusion models

| Parameters | $R^2$ | RMSE |
|---|---|---|
| Jmax | 0.76 | 8.84 |
| Vcmax | 0.71 | 7.84 |

In addition, decision fusion often means that the final predicted values of different samples are obtained in different ways. The decision-making logic in the process provided a basis for researchers to analyze the mechanisms of different models.

## 4 Discussion

**4.1 Connections with previous studies**

Although there have been many studies exploring vegetation photosynthetic capacity prediction models and accumulating huge data sets, few studies have simultaneously measured SIF and reflectance at the same location. In this study, through the results of the single data sources and the data fusion models, it can be found that SIF had a better prediction effect on Jmax, while reflectance has a better prediction effect on Vcmax. Although there was currently a lack of public data sets comparing the two at the same time, some studies have supported the conclusions of this study. Barnes et al. (2017) used PLSR and hyperspectral indices to predict photosynthetic capacity and found that the results for Vcmax were generally better than those for Jmax. At the same time, past studies have also found that the robustness of different data sources to different photosynthetic capacity parameters was greatly different. Watt et al. (2020) found that SIF had a very high correlation with Jmax and Vcmax. However, when only considering samples under P limiting conditions, the correlation between SIF and Vcmax was greatly reduced, while the correlation between SIF and Jmax did not change much. Research by Meacham-Hensold et al. (2019) showed that when the reflectance-based prediction model involved many years, the prediction effect for Jmax would drop a lot, but the impact on Vcmax would be limited.

Especially since most methods for estimating SIF involve Fraunhofer lines, often only values at the hydrogen (H, 656.4 nm) absorption band and the two oxygen ($O_2$-A, 760.4 nm; $O_2$-B, 687.0 nm) absorption bands are obtained (Meroni et al. 2009). Currently there were only a few studies that evaluate the chlorophyll content, nitrogen, etc. of plant leaves based on complete SIF or SIF yield spectra (Xu et al. 2023; Yin et al. 2023). There were no studies that used complete SIF or SIF yield spectra to evaluate the photosynthetic capacity of leaves. Ding et al. (2023) presented a semimechanistic model to estimate the seasonal Vc,max and stomatal conductance based on the field SIF observations. However, this model required some prior knowledge and did not evaluate another important parameter Jmax. Shi et al. (2022) used an improved MLR-SIF model to capture the net photosynthetic rate of trees. But this method required a long period of dark adaption for each measurement. Compared with other SIF-based methods, the prediction models proposed in this study had faster measurement speed and did not require tedious pre-experiments. In addition, the feature extraction results also provide a theoretical basis for the production of miniaturized instruments.

**4.2 Mechanisms of SIF-based photosynthesis trait estimation**

Based on the latest radiation transfer model PROSPECT-PRO, the reflectance of 400-2500 nm was input to obtain the CBC and proteins of the rice leaves samples. After removing the abnormally small simulated values of few samples, the correlations of reflectance, upward SIF yield and downward SIF yield per band between *Vc,max*, *Jmax*, Proteins and CBC were obtained, respectively. According to different data sources, the results were drawn as correlation line charts as Figure 6.

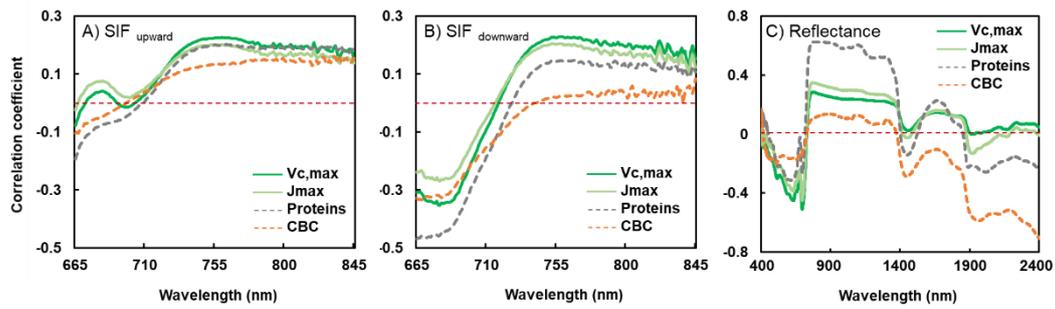

**Fig. 6.** The correlation coefficient of reflectance, upward SIF yield and downward SIF yield per band between $V_{c,max}$, $J_{max}$, Proteins and CBC. The dark green and light green solid lines represent $V_{c,max}$ and $J_{max}$, respectively, and the gray and orange dotted lines represent proteins and CBC, respectively. The dashed line parallel to the X-axis had a Y-axis value of 0.

Similar to many previous studies on reflectance and nitrogen, there is a strong correlation in the visible and near-infrared bands (Feng et al. 2008; Zhao et al. 2005). The per-band correlation coefficients of SIF spectra and nitrogen were similar to previous studies at moderate nitrogen levels (Jia et al. 2018). It can be found that the SIF downward and nitrogen before 680 nm have a high correlation. In addition, it can also be found from Figure 3 that the correlation curve between leaf reflectance and photosynthetic capacity is very similar to the correlation curve between reflectance and nitrogen, and both have peaks and valleys at the same wavelength. Therefore, some researchers thought the $V_{c,max}$ and $J_{max}$ estimations from leaf reflectance are predominantly based on their relationships to leaf nitrogen (Dechant et al. 2017). As a unique remote sensing signal stimulated by chlorophyll, as reported by a significant amount of literatures, there is a strong correlation between chlorophyll and leaf nitrogen (Baret et al. 2007; Clevers and Kooistra 2012), so the correlation curves between SIF and photosynthetic capacity, and SIF and leaf nitrogen are inevitably similar. This performance is more obvious on SIF downward (Figure 3.B). however, it was worth noting that the correlation between SIF and plant photosynthetic capacity was greater than the correlation between SIF and biochemical parameters in many bands, which was completely different from reflectance.

The VIP algorithm helps explain variable importance in PLSR and has been widely used in different fields (Farrés et al. 2015; Wold et al. 1993). The results are shown in the Figure 7.

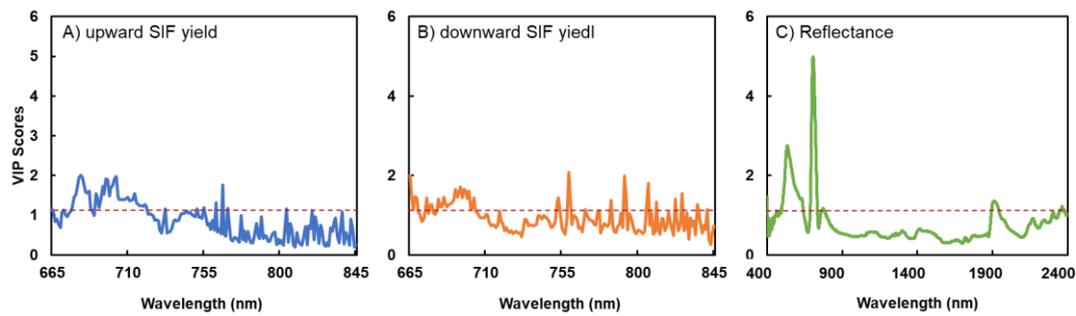

**Fig. 7.** Variable importance in projection (VIP) scores' results of the partial least-squares (PLS) regression models for the three data sources. Variables with a VIP score greater than 1 are considered important for the projection of the PLSR models.

It can also be seen from the importance of each band to the PLSR models that the visible bands related to pigments are crucial to the reflectance-based model. However, existing research showed that the correlation between chlorophyll and reflectance often did not apply in the case of high chlorophyll content (Van Wittenberghe et al. 2014). From the Figure C1 in Appendix C, when the chlorophyll content was greater than 80 μg cm$^{-2}$, the reflectance-based model was difficult to evaluate *Jmax* and *Vc,max* (both $R^2$ were less than 0.05), but SIF can still make an effective evaluation, among which upward SIF yield-based model's $R^2$ can even reach about 0.5. At mid and low chlorophyll contents, the difference between the three data sources were not so obvious. In addition to the band before 700 nm in the SIF-based model, several bands after 750 nm also contribute to the model. The bands that have a significant relationship with chlorophyll make a large contribution to the reflectance-based model. The biochemical indicators corresponding to each band of reflectance have been recognized by researchers, while the physical and biological meanings represented by the SIF spectra of different bands need to be further explored. However, the above analysis shows that the mechanism of the SIF-based model may be different from the reflectance-based model for evaluating photosynthetic capacity.

**4.3 Application of Data Fusion Strategies**

Leaf nitrogen is involved in many leaf physiological processes, in addition to photosynthesis, respiration, structural growth and storage capacity building (Liu et al. 2018). And reflectance may have a better nitrogen assessment effect than SIF. But at the same time, the close relationship between SIF and plant physiological state makes it irreplaceable. Multiple N-based leaf biochemical constituents tend to change throughout the plant leaf cycle in response to changing environmental factors (Kokaly et al. 2009). Therefore, to a certain extent, SIF and reflectance spectra represent different physiological and biochemical characteristics of leaves, which is also proved by the successful practice of the data fusion models in chapter three.

The measurement-level data fusion may have some limitations which are a high data volume and the possible predominance of one data source over the others (Borras et al. 2015). Feature-level fusion requires less input to facilitate the development of real-time systems, and some researchers have developed related methods based on the results of feature-level data fusion (Adeel et al. 2019; Moshou et al. 2005). Even if the decision-level data fusion did not bring about a particularly obvious improvement in performance, it could better help us understand the mechanism of different data sources for evaluating different photosynthetic parameters. Because in the decision-level data fusion for modeling, only reflectance data was selected on some occasions, and only SIF was selected on some occasions, and three data sources were selected on some occasions.

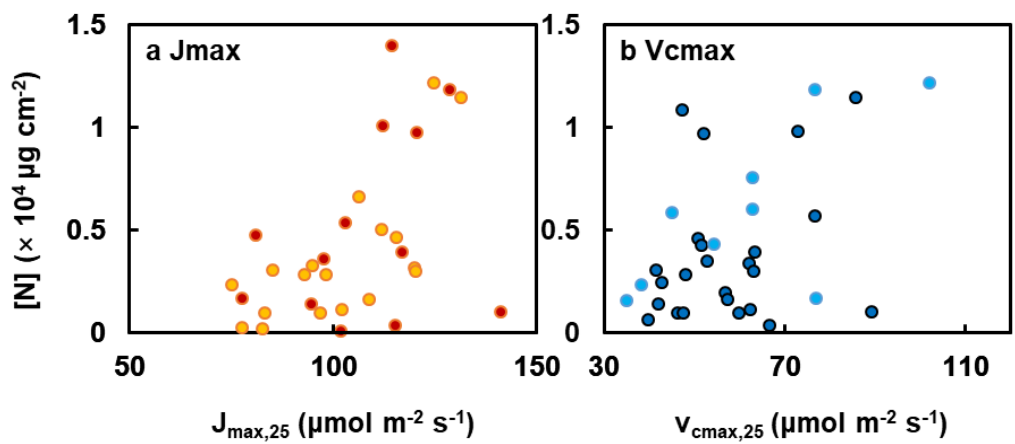

**Fig. 8.** Nitrogen distribution of measured values in decision-level data fusion results. Light colored points indicated that the data used the average of the three data sources' prediction results; dark colored points indicated that the data used the median value of the three data sources.

It can be seen from the Figure 8 that in the prediction of *Jmax*, the "mean value" strategy was used in more cases, while in the prediction of *Vc,max*, the "intermediate value" strategy was more common. At the same time, when the leaf chlorophyll content was high, the predicted value often only involved SIF, which was consistent with the results of independent modeling of each data set. The sample size of the data set used in this study is still limited compared to more mature models for chlorophyll, nitrogen and other chemical indicators, and the information increase brought by data fusion may not be able to make up for this deficiency. In the future, more data may be needed to improve the accuracy of the model (Biancolillo et al. 2014).

**4.4 Potential applications and future perspectives**

Currently, global scale plant photosynthetic capacity maps can be drawn based on SIF or other remote sensing technology (He et al. 2019; Luo et al. 2021). Different modeling methods can even lead to wildly different predictions for the same area. Reflectance has inherent shortcomings of the data itself such as delayed sensitivity and lack of specificity, which are difficult to solve through data processing and other means (Tremblay et al. 2011; Van Wittenberghe et al. 2014). The SIF may be affected by satellite viewing angle, instrument accuracy and vegetation structure. Data fusion strategies may be able to more accurately assess outcomes in these locations.

On the other hand, in the future, it may be considered to combine the radiative transfer model to invert the canopy SIF spectra to make up for the shortcoming of only measuring the SIF value at the Fraunhofer line and quickly predict the plant photosynthetic capacity at the canopy scale.

# 5 Conclusion

The results of this study showed that the PLSR models based on reflectance and SIF spectra can evaluate the photosynthetic capacity of crops. However, compared to the indirect assessment of reflectance based on nitrogen, SIF can directly assess photosynthetic capacity, and is far superior to the effect of reflectance in the case of high chlorophyll content. Various levels of data fusion strategies can improve the usability of the models at different levels. Based on the data fusion strategies combined with the feature extraction algorithm, it helped to improve the accuracy of the models while reducing the amount of input. The successful practice of plant phenotyping technology in evaluating the photosynthetic capacity of leaves has shortened the measurement time by more than 10 times compared with the traditional gas exchange method, and many conclusions have the potential to be promoted to the canopy or even satellite scale. The main limitation of this study is that the dataset for the models is mainly from a specific $C_3$ crop rice. Therefore, the relationship of spectra to parameters of photosynthetic capacity, such as the exact wavelengths identified here, may not guarantee accuracy when applied to other $C_4$ crops such as maize under different conditions. Further studies on more crop or vegetation species for different photosynthetic capacity parameters and at different scales will be important to evaluate and enrich the methods developed in this paper.

# Appendix

Appendix A

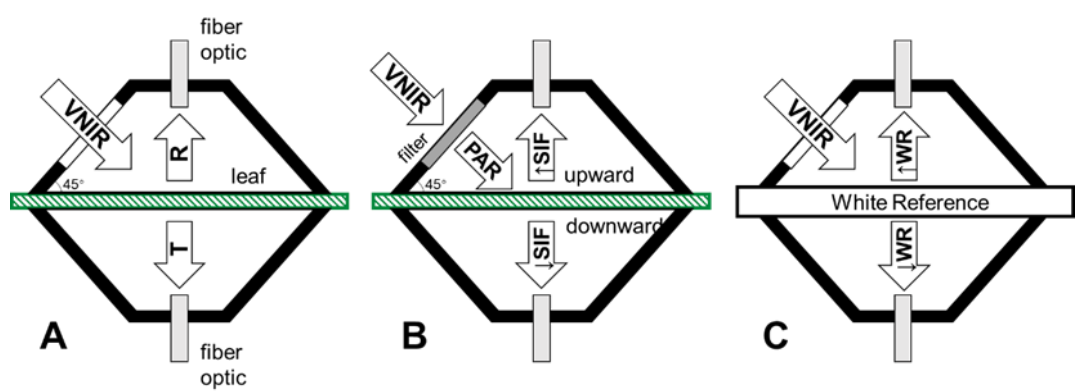

Figure A. Schematic of the FluoWat leaf clip. Reflectance (R), transmittance (T) and sun-induced fluorescence (SIF) are measured in the visible light (VIS) and near infrared (NIR) wavelength ranges (350 – 2,500 nm) by placing a fiber optic either in the upward (↑) or downward (↓) position (subgraph A). The inner surface of the clip is coated with mate black synthetic paint (3% flat reflectance over the whole solar range). After placing the shortpass to restrict incoming PAR to visible wavelengths up to 650 nm (subgraph B), upward and downward sun-induced fluorescence (↑SIF, ↓SIF) are measured. The total light arriving to the sample is obtained by measuring with the white reference, and then PAR absorbed by the sample is calculated through the R and T obtained from the subgraph A, and finally the SIF yield is obtained (subgraph C).

Appendix B

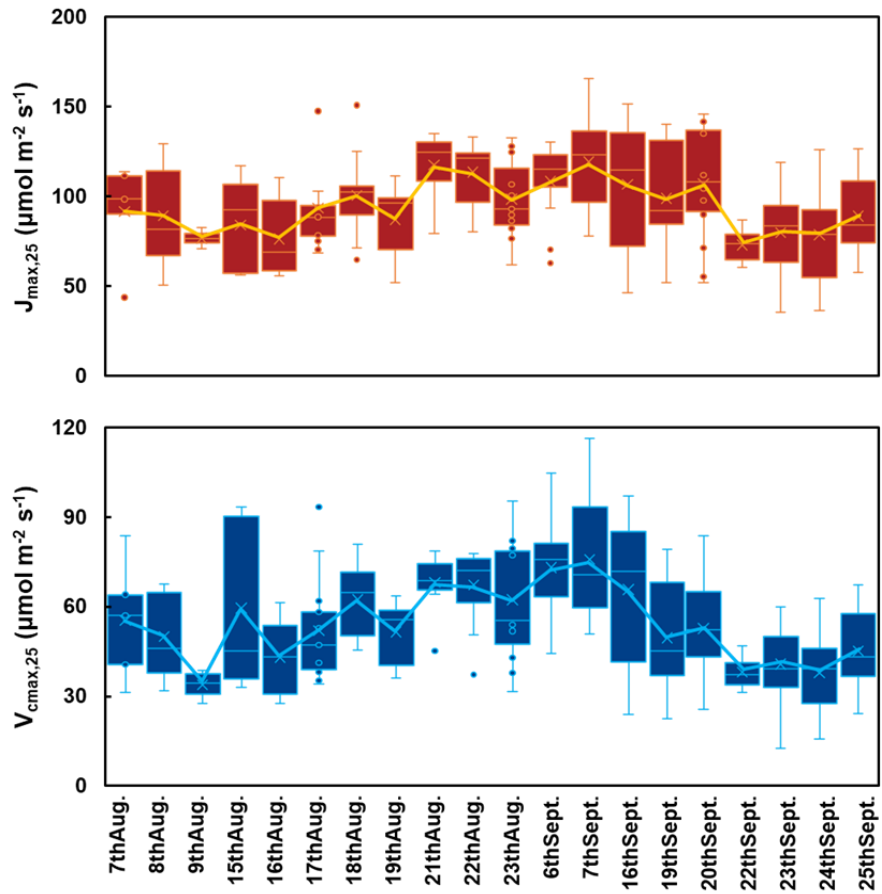

**Figure B1.** Summary statistics of the leaf photosynthetic capability parameters calculated by *A/Ci* curves.

**Table B2.** Summary statistics of leaf traits measured in the net room. Mean, standard deviation, minimum, and maximum values are given from maximum electron transport rate (*Jmax*) and maximum carboxylation capacity (*Vc,max*) from *A/Ci* curves.

| Trait | Status | Mean | Stand. Dev. | Min. | Max. |
|---|---|---|---|---|---|
| Jmax | $N_1$ | 94.45 | 13.91 | 69.90 | 117.84 |
|  | $N_2$ | 94.68 | 14.24 | 57.86 | 126.43 |
|  | $N_3$ | 101.95 | 16.61 | 69.63 | 141.17 |
|  | HHZ | 98.60 | 12.90 | 69.90 | 117.84 |
|  | XS134 | 96.74 | 15.21 | 57.86 | 141.17 |
| Vc,max | $N_1$ | 55.74 | 10.68 | 35.17 | 72.79 |
|  | $N_2$ | 53.04 | 12.63 | 29.11 | 91.14 |
|  | $N_3$ | 59.70 | 15.28 | 29.26 | 101.93 |
|  | HHZ | 57.82 | 10.65 | 29.26 | 76.83 |
|  | XS134 | 55.27 | 15.21 | 29.11 | 101.93 |

Appendix C

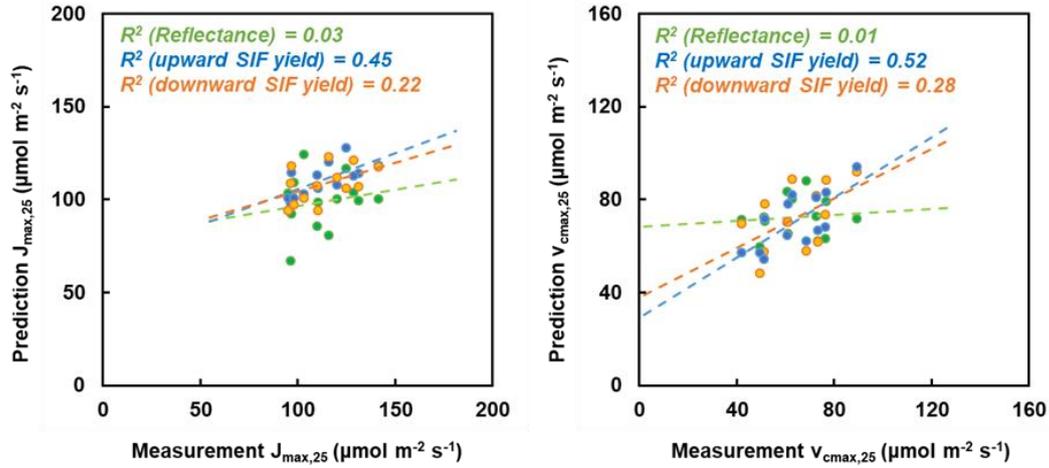

Figure C1. PLSR prediction models built based on high chlorophyll content (> 80 μg cm$^{-2}$). Green, blue and orange points represented reflectance, upward SIF yield and downward SIF yield, respectively. The dotted lines represented the trend lines. The coefficient of determination was shown in the figure.

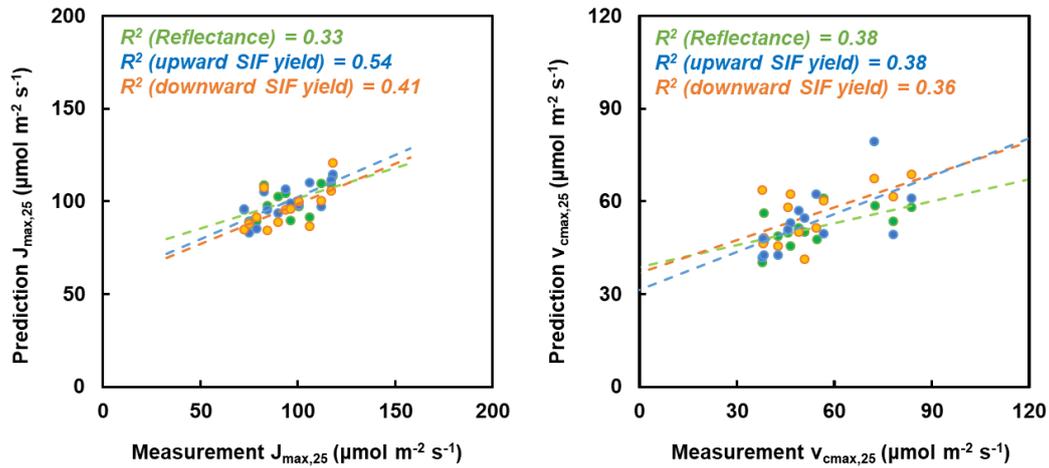

Figure C2. PLSR prediction models built based on mid chlorophyll content (60-80 μg cm$^{-2}$). Green, blue and orange points represented reflectance, upward SIF yield and downward SIF yield, respectively. The dotted lines represented the trend lines. The coefficient of determination was shown in the figure.

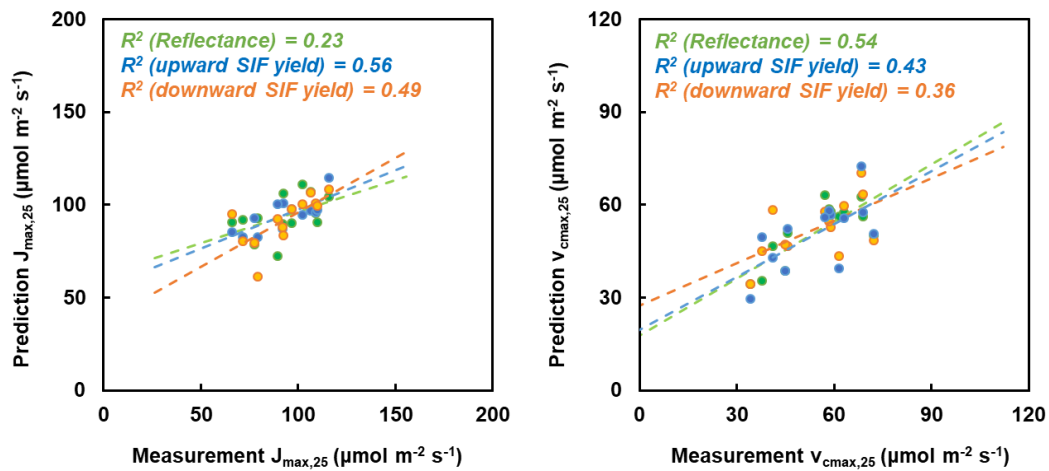

Figure C3. PLSR prediction models built based on low chlorophyll content (<60 µg cm$^{-2}$). Green, blue and orange points represented reflectance, upward SIF yield and downward SIF yield, respectively. The dotted lines represented the trend lines. The coefficient of determination was shown in the figure.